\documentclass[aps,prb,twocolumn,showpacs,superscriptaddress]{revtex4-1}
\usepackage{amsfonts}
\usepackage{amsmath}
\usepackage{amssymb}
\usepackage{graphicx}
\usepackage{float}
\usepackage{multirow}
\usepackage{ulem}
\usepackage{gensymb}
\usepackage{dcolumn}
\usepackage[dvipsnames]{xcolor}

\usepackage[bookmarks=false]{hyperref} 
\hypersetup{colorlinks=true,
            linkcolor=blue,
            citecolor=blue,
            urlcolor=orange}
\begin{document}
\title{DFT+DMFT study of exchange interactions in cobalt and their implications \\for the competition of hcp and fcc phases}
\author{A. A. Katanin}
\affiliation{Center for Photonics and 2D Materials, Moscow Institute of Physics and Technology, Institutsky lane 9, Dolgoprudny, 141700, Moscow Region, Russia}
\affiliation{Skolkovo Institute of Science and Technology, 121205 Moscow, Russia}
\affiliation{M. N. Mikheev Institute of Metal Physics of Ural Branch of Russian Academy of Sciences, S. Kovalevskaya Street 18, 620990 Yekaterinburg, Russia}

\begin{abstract}
We reconsider magnetic properties of fcc and hcp cobalt within the density functional theory plus dynamical mean-field theory (DFT+DMFT) approach in the paramagnetic phase. Using recently proposed approach of calculation of exchange interactions in paramagnetic phase, we extract exchange interaction parameters of fcc and hcp cobalt and show that the hcp phase possesses larger spin stiffness, in agreement with the experimental data, showing stronger tendency to ferromagnetism. Accordingly, the DMFT Curie temperature of the hcp phase appears to be higher, than that of the fcc phase. The disappearance of magnetic order in the fcc phase well below the cobalt Curie temperature is expected to affect its structural stability and make this phase energetically unfavourable near the experimental Curie temperature. This  may explain the ``revival" of the hcp phase near the Curie temperature of cobalt, observed in the recent experimental results of perturbed angular correlation study [Sci. Rep. {\bf 12}, 10054  (2022)].
\end{abstract}. 
\maketitle
\section{Introduction} 
Magnetism of transition metals remains a cornerstone of the theory of magnetism, since on one hand they are often described by Heisenberg theory, but on the other hand they remain itinerant, showing fractional magnetic moments \cite{Goodenough,Vonsovsky,Mattis,Moriya}. While elemental iron represents the case of strong well localized magnets, elemental nickel appears to be much more itinerant, due to substantial deviation from half filling. 

In this respect, cobalt, having magnetic moment only slightly smaller than iron, and the Curie temperature $T_C\simeq 1400$~K, which is even higher than in iron ($\simeq 1000$~K), is interesting for both, theoretical and experimental research. Understanding of magnetic properties of cobalt is complicated by the structural hcp-fcc transition\cite{PhaseTr1,PhaseTr} at $T_{\rm hcp-fcc}\simeq 700$~K. The energies of these phases are sufficiently close, such that  small external perturbations easily shift this temperature \cite{Mutual1,Mutual2,Mutual3}. Recent perturbed angular correlation study \cite{PAC} has suggested that in fact, substantial amount of fcc phase appears only at $T\sim 500$~K, and at higher temperatures $T\gtrsim 700$~K the hcp phase reappears, although with suppressed magnetization. Quite interestingly, similar observations, although at somewhat shifted temperatures, can be also made from the old measurements of the molar volume (see Ref. \onlinecite{CoV} for a review), which increases by $\Delta V_m \sim 2\cdot 10^{-8} m^3/mol$ at $T\sim 400$~K and at higher temperatures $T\gtrsim 1200$~K two different sets of measurements lead to the  molar volumes different approximately by $\Delta V_m$. These changes of the molar volume may indicate possible phase transformations (or mixture of different phases), which require theoretical explanations. The decrease of the content of the fcc phase at temperatures close to Curie temperature was also observed in Ref. \onlinecite{fccDecrease}. For the transition from the low-temperature hcp to the higher temperature fcc phase, it was suggested that magnetic correlations are important, possibly in combination with the other (e.g., phonon) contributions, see, e.g., Refs. \onlinecite{SpinFluct,EnergiesNew} and references therein. Studying the stability of fcc and hcp phases of cobalt in the vicinity of Curie temperature represents an important theoretical problem.

Both, hcp and fcc phases of cobalt were studied\cite{DFT00, DFT01, DFT0,DFT1,DFT2,DFT3,DFTJ,DFT4,DFT5} by the density functional theory (DFT), which however does not include correlation effects and meets difficulties in describing the effect of temperature fluctuations. The combination of DFT with the dynamical mean-field theory (DMFT) method \cite{DMFT_rev,DFTplusDMFT} was recently applied to the fcc cobalt \cite{BelozerovCo}, yielding, however, the Curie temperature, which is smaller than the experimental one, while typically the mean-field approaches overestimate magnetic transition temperature. Therefore, the magnetic order in the fcc phase can be destroyed below the experimental Curie temperature of cobalt. Since presence of long-range magnetic order decreases the energy of fcc phase by $\sim 0.2$~eV, which is an order of magnitude larger than the energy difference of fcc and hcp phases in ferromagnetic state \cite{EnergiesNew}, the fcc phase may become also thermodynamically unstable due to loss of the long-range magnetic order.  

In the present paper we revisit the properties of hcp and fcc phases of cobalt within the DFT+DMFT approach. We use the lattice parameters at the experimental Curie temperature of cobalt to account for thermal expansion of the lattice and calculate more accurately magnetic transition temperatures. We furthermore apply the recently proposed technique of calculation of exchange interactions in paramagnetic phase within the DFT+DMFT approach \cite{MyJ}. Consideration of the exchange interaction in the symmetric phase assumes the existence of well-defined local magnetic moments and provides a possibility of unbiased evaluation of exchange interactions, not affected by the considered type of magnetic order. Using the obtained exchange interactions, we also estimate the non-local corrections to the DMFT Curie temperatures and energies of fcc and hcp phases.

On one hand, we show that using the high-temperature lattice parameter yields larger Curie temperature in DMFT, which is closer the experimental data. Yet, the non-local correlations decrease the Curie temperature of the fcc phase to $T_C\simeq 750$~K, which is much smaller than the experimental value, but surprisingly close to the temperature of the experimental observation of the ``revival" of hcp phase. The Curie temperature of the hcp phase with account of non-local correlations appears $T_C\simeq 1300$~K, quite close to the experimental data \cite{PAC}.

Therefore, on the basis of these results, we suggest the following physical picture, which agrees with recent experimental data. The hcp phase of cobalt is stable at low temperatures, being lower in the energy in ferromagnetic state, than the hcp phase. With increase of temperature the fcc phase decreases its energy due to vibrational, magnetic, and electronic degrees of freedom (see Ref. \onlinecite{EnergiesNew}), which first yields that this phase becomes energetically preferable. But this phase becomes paramagnetic at $T\sim 700$~K, which again favours the hcp phase. Therefore, we expect delicate intertwinning of magnetic and structural properties in cobalt.   

{The plan of the paper is the following. In Sect. II, we discuss used methods, in particular the method of calculation of the magnetic exchange interaction from the inverse momentum-dependent susceptibility. In Sect. III, we describe the results of the DFT+DMFT approach. In Sect. IV we present our conclusions.}  

\section{Methods}

The DFT calculations of cobalt were performed using the pseudo-potential method implemented in the Quantum Espresso \cite{QE} package supplemented by the maximally localized Wannier projection onto $3d$, $4s$, $4p$ states performed within Wannier90 package \cite{Wannier90}. To take into account the effect of thermal expansion on Curie temperatures, we use the experimental volume of the unit cell of fcc cobalt at the Curie temperature $V=11.66~$\AA$^3$ (Ref. \onlinecite{CoV}), corresponding to the lattice parameter $a=3.60$~\AA, see also Ref. \onlinecite{PhaseTr}. For comparison, we also perform some calculations for the lattice parameter $a_0=3.56$~\AA, accepted in Ref. \onlinecite{BelozerovCo}, which is closer to the zero-temperature lattice constant.
For hcp cobalt we choose the ratio $c/a=1.63$ corresponding to the temperatures $T\sim T_C$. Taking into account that the experimental unit cell volume per atom quite weakly changes at the fcc-hcp transition and fixing the volume equal to that of the fcc phase, we put $a=2.55$~\AA.  
The reciprocal space integration was performed using ${16\times 16\times 16}$\, ${\bf k}$-point grid for fcc phase and ${18\times 18\times 18}$\, ${\bf k}$-point grid for hcp phase. 

In DMFT calculation we consider the density-density interaction matrix, parameterized by Slater parameters $F^0$, $F^2$, and $F^4$, expressed through Hubbard $U$ and Hund $J_H$ interaction parameters according to 
${F^0\equiv U}$ and ${(F^2+F^4)/14 \equiv J_{\rm H}}$, $F_2/F_4\simeq 0.63$ (see Ref.~\onlinecite{u_and_j}). In the present work we take $U=4$~eV, $J_H=0.9$~eV. We use a double-counting correction ${H}_{\rm DC} = M_{\rm DC}\sum_{ir}  {n}_{ird} $ in the around mean-field form~\cite{AMF}, $M_{\rm DC}=\langle {n}_{ird} \rangle [U (2 n_{\rm orb} {-} 1) - J_{\rm H}  (n_{\rm orb} {-} 1)] / (2 n_{\rm orb})$, where
${n}_{ird}$ is the {operator of the} number of $d$ electrons at the site $(i,r)$, where $i$ is the unit cell index and $r$ is the site index within the unit cell. For the low-symmetry hcp phase we perform the Hamiltonian rotation in the d-orbital space to diagonalize the crystal field, which considerably reduces the off-diagonal components of the local Green's functions with respect to the orbital indexes and improves applicability of the density-density interaction.

We define the exchange interaction by considering the effective Heisenberg model with the Hamiltonian $H=-(1/2)\sum_{{\bf q},rr'} J^{rr'}_{\bf q} {\mathbf S}^r_{\mathbf q} {\mathbf S}^{r'}_{-{\mathbf q}}$, {$\mathbf S^r_{\mathbf q}$ is the Fourier transform of static operators ${\mathbf S}_{ir}$},
where the orbital-summed on-site static spin operators ${\mathbf S}_{ir}=\sum_m {\mathbf S}_{irm}$ and ${\mathbf S}_{irm}=(1/2)\sum_{\sigma\sigma'\nu}c^+_{irm\sigma\nu}\mbox {\boldmath $\sigma $}_{\sigma\sigma'}c_{irm\sigma'\nu}$  
is the electron spin operator, $\nu$ are the Matsubara frequencies, $c^+_{irm\sigma\nu}$ and $c_{irm\sigma\nu}$ are the frequency components of the electron creation and destruction operators at the site $(i,r)$, $d$-orbital $m$, and spin projection $\sigma$, and $\mbox {\boldmath $\sigma $}_{\sigma \sigma'
}$ are the Pauli matrices.

To extract the exchange parameters $J_{\bf q}$, we relate them to the orbital-summed non-local static {longitudinal} susceptibility $\chi^{rr'}_{\mathbf q}=-\langle \langle S^{z,r}_{\mathbf q}|S^{z,r'}_{-{\mathbf q}}\rangle\rangle_{\omega=0}=\sum_{mm^{\prime}}\hat{\chi}_{\bf q}^{mr, m^{\prime}r'}$ (the hats stand for matrices with respect to orbital and site indexes; $\langle \langle ..|..\rangle\rangle_\omega$ is the retarded Green's function), considering the generalization of the approach of Ref. \onlinecite{MyJ} to several atoms in the unit cell, and express exchange interactions as 
\begin{equation}
J_{\mathbf q}=
\chi_{\rm loc}^{-1}-\chi_{\bf q}^{-1},
\label{JqAvDef}
\end{equation}
 the inverse in Eq. (\ref{JqAvDef}) is taken with respect to the site indexes in the unit cell. The matrix of local susceptibilities $\chi^{rr'}_{\rm loc}=-\langle \langle S^z_{ir}|S^z_{ir}\rangle\rangle_{\omega=0}\delta_{rr'}=\sum_{{  m}{  m}'} \hat{\chi}^{{  m}{  m}',r}_{\rm loc}\delta_{rr'}$ is diagonal with respect to the site indexes. For the non-local susceptibility we use (cf. Refs.~\onlinecite{MyJ,MyGamma,MyEDMFT,OurRev,EdwHrtz})
\begin{equation}
\hat{\chi}^{}_{\bf q}=\frac{1}{2}\left[\hat{\Pi}_{\bf q}^{-1}-\hat{U}^{s}\right]^{-1}, \label{phi}
\end{equation}
with the static spin polarization {(irreducible static spin susceptibility)} $\hat{\Pi}_{\bf q}$ (see Ref. \onlinecite{MyJ} for the procedure of its evaluation) and  $\hat{U}^s=\hat{U}_{\uparrow\downarrow}-\hat{U}_{\uparrow\uparrow}$ is the electron interaction matrix in the spin channel, and the matrix inversions in Eq. (\ref{phi}) are assumed.

The alternative approach considered in Ref. \onlinecite{MyJ} uses the inverse of orbital-resolved susceptibilities, $J^{mr,m'r'}_{\mathbf q}=(\hat{\chi}_{\rm loc}^{m m^{\prime},r})^{-1}\delta_{rr'}-(\hat{\chi}_{\bf q}^{mr, m^{\prime}r'})^{-1}$ and averages the respective orbital-resolved exchange interactions over orbitals with the local susceptibilities, $J_{\mathbf q}^{rr}=\sum_{mm'} J^{mr,m'r}_{\mathbf q} \chi_{\rm loc}^{mm',r}/\sum_{mm'}\chi_{\rm loc}^{mm',r}$. For many atoms in the unit cell this approach can be applied to the diagonal components of magnetic exchange $J^{rr}_{\mathbf q}$ only, since the local susceptibilities are diagonal with respect to the site index. As it is discussed in Ref. \onlinecite{MyJ} the results of this approach are expected to be close to those from the Eq. (\ref{JqAvDef}) if the local magnetic moments are well formed. We therefore analyze the results of this alternative approach too for comparison purposes.

The DMFT calculations of the self-energies, non-uniform susceptibilities and exchange interactions were performed within the Wan2mb software package \cite{MyJ,WanMb}, based on the continuous-time Quantum Monte Carlo (CT-QMC) method of the solution of impurity problem\cite{CT-QMC}, realized in the iQIST software \cite{iQIST}. In the calculation of vertices we account for $60$-$90$ fermionic frequencies (both positive and negative). In the summations over frequencies, the corrections on the finite size of the frequency box are accounted according to Refs.~\onlinecite{MyEDMFT,MyJ}. 


\section{DFT+DMFT calculation results}

\subsection{Local spectral functions}

\begin{figure}[b]
		\center{		\includegraphics[width=1.\linewidth]{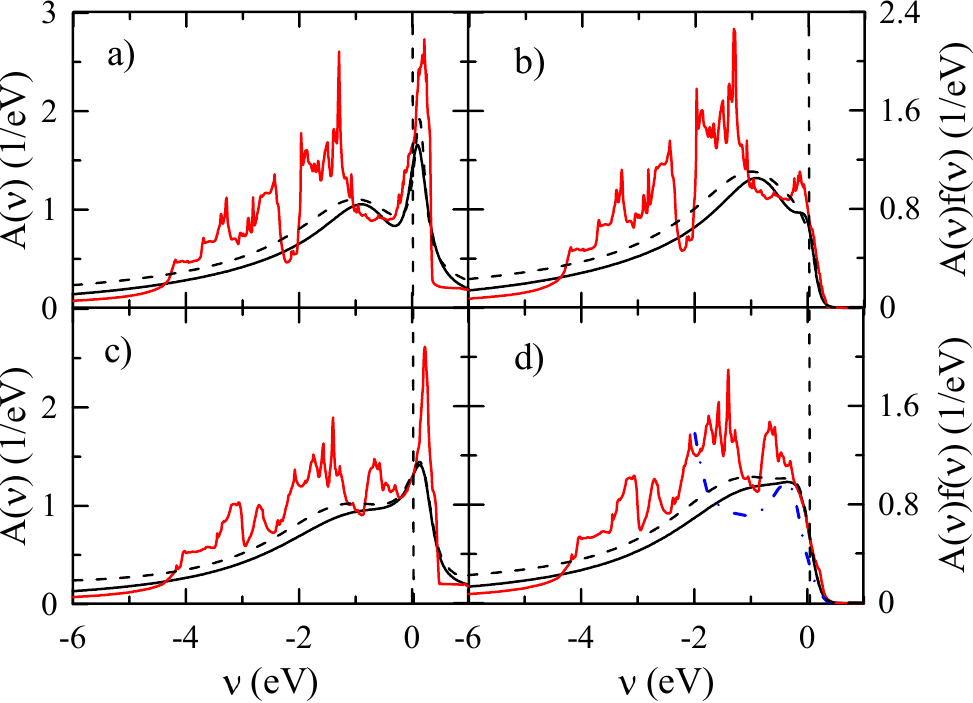}}
		\caption{
Energy dependence of the momentum-integrated spectral functions (densities of states) $A(\nu)$ (a,c) and those weighted by the Fermi function, $A(\nu) f(\nu)$ (b,d) in the hcp (a,b) and fcc (c,d) phases. Black dashed (solid) lines correspond to the total (partial for $d$ electrons) $A(\nu)$ in DFT+DMFT approach at $\beta=10$~eV$^{-1}$; red lines correspond to the DFT approach. Blue dot-dashed line in (d) shows the photoemission experimental data  in the fcc phase of thin cobalt film \cite{PES_exp2} (taken at the energy $\hbar \omega=16$~eV at $T=350$~K$=1.1 T_C^{\rm film}$). Vertical dashed lines mark the position of the Fermi level.}
\label{Fig_Aw}
\end{figure}

In Fig. \ref{Fig_Aw} we show the DMFT local densities of states $A(\nu)$ at $\beta=10$~eV$^{-1}$, together with $A(\nu)$ weighted by the Fermi function $f(\nu)$ at the same temperature. The $d$-states provide major contribution to the density of states near the Fermi level. For comparison we also show the corresponding DFT densities of states (together with those weighted by $f(\nu)$). One can see that in both, DFT and DFT+DMFT approaches the hcp cobalt has somewhat stronger peak of the density of states near the Fermi level (see also Ref. \onlinecite{DFT00} for comparison of the DFT densities of states). The interaction effects lead to suppression of the peaks of the DFT density of states below the Fermi level, yielding formation of the maximum at $\nu\simeq -1$~eV in the hcp phase and the plateau at $\nu\lesssim -0.4$~eV in the fcc phase. 

For fcc phase we compare the resulting spectral function with the available experimental data on thin cobalt film above the corresponding Curie temperature \cite{PES_exp2}. For comparison purposes we scale the experimental spectral function (expressed in arbitrary units) to the height of the maximum of the DFT+DMFT spectral function. The resulting dependence near the Fermi level and the position of the plateau of the density of states are reproduced in DFT+DMFT approach (as it is discussed in Ref. \onlinecite{PES_exp2} the increase of the experimental spectral function at $\nu<-2$~eV is due to cupper substrate). 


\subsection{Local and uniform susceptibilities}

\begin{figure}[b]
		\center{		\includegraphics[width=0.65\linewidth]{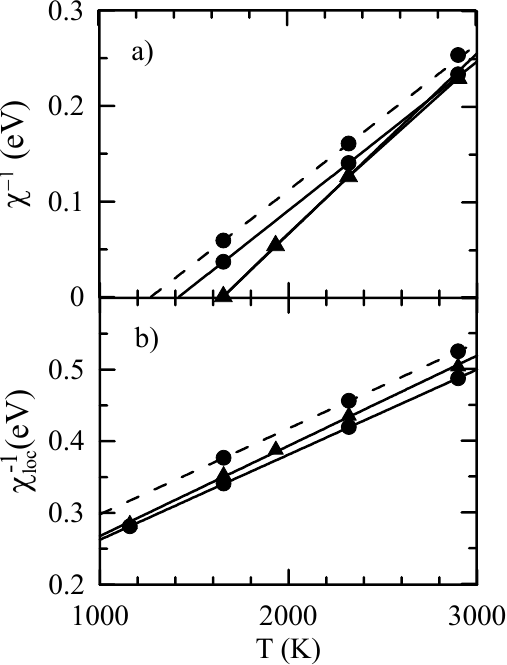}}
		\caption{
Temperature dependence of the inverse uniform (a) and local (b) susceptibilities of cobalt within the   DFT+DMFT approach. Solid lines with circles correspond to the fcc lattice with the lattice parameter $a$ at $T=T_C\simeq 1400$~K, dashed lines show the result for the fcc lattice with the low-temperature lattice parameter $a_0$, considered in Ref. \onlinecite{BelozerovCo}, solid line with triangles corresponds to the hcp lattice. }
\label{Fig_chi0_Co}
\end{figure}


The temperature dependence of the inverse uniform susceptibility in fcc and hcp phases is shown in Fig. \ref{Fig_chi0_Co} (for the momentum dependencies see Appendix). For the chosen lattice parameter $a$ of the fcc phase, the inverse susceptibility for fcc structure vanishes at the Curie temperature $T^{\rm DMFT}_{C,{\rm fcc}}=1420$~K. At the same time, with the low-temperature lattice parameter $a_0$ we obtain lower $T^{{\rm DMFT},0}_{C,{\rm fcc}}\simeq 1270$~K, which is close to the result of Ref. \onlinecite{BelozerovCo} and remains lower than the experimental Curie temperature. The decrease of the spin susceptibility and Curie temperature with decrease of the lattice parameter is explained by weakening correlation effects due to increase of hopping parameters, which make the system more itinerant. For hcp phase we obtain larger DMFT Curie temperature $T^{\rm DMFT}_{C,{\rm hcp}}=1620$~K. 
Due to the mean-field nature the dynamical men-field theory approach is known to overestimate Curie temperature. Therefore, obtained Curie temperatures can be considered as an upper bound and corrected below with account of the non-local correlations.

From the slope of inverse magnetic susceptibilities extract the local magnetic moment. In fcc phase with the lattice parameter $a=3.60$~\AA~we find $\mu_{\rm loc}^2=8.7\mu_B^2$, in terms of the effective spin, defined by $g^2 p_{\rm loc}(p_{\rm loc}+1)=\mu_{\rm loc}^2$ ($g=2$), this corresponds to $p_{\rm loc}=1.06$. From the uniform susceptibility we obtain somewhat smaller magnetic moment $\mu_{\rm loc}^2=6.6\mu_B^2$. The Weiss temperature $T_W$ of the inverse local magnetic susceptibility 
\begin{equation}
\chi_{\rm loc}^{-1}=3(g\mu_B)^2 (T+T_W)/\mu_{\rm loc}^2
\label{chiloc}
\end{equation}
determines the Kondo temperature\cite{Wilson,Wilson1,Melnikov,Tsvelik,MyComment} $T_K\approx T_W/\sqrt{2}\simeq 860$~K, which is rather large. For the hcp phase we obtain somewhat smaller magnetic moment $\mu_{\rm loc}^2=8.22\mu_B^2$, corresponding to $p_{\rm loc}\simeq 1$, the respective Kondo temperature $T_K\simeq 800$~K. The uniform susceptibility in this phase yields $\mu^2=5.47\mu_B^2$. The obtained Kondo temperatures are comparable to those in nickel \cite{MyJ,Sangiovanni}. However, the magnetic moment in cobalt is relatively well formed \cite{BelozerovCo},
and large Kondo temperature originates from its larger value $p_{\rm loc}\approx 1$,  providing at the same time more channels to the screening of this local magnetic moment. 

\subsection{Exchange interactions}

 \begin{figure}[t]
		\center{	
\includegraphics[width=0.95\linewidth]{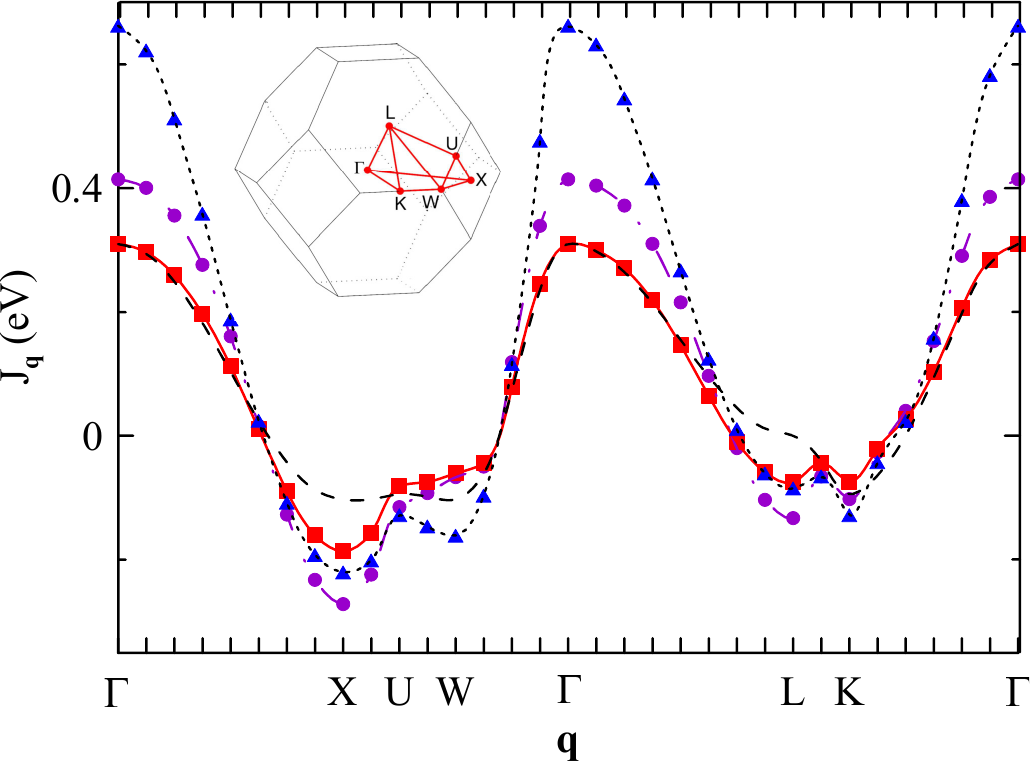}}
		\caption{Momentum dependence of the exchange interactions in the fcc phase of cobalt at $\beta=7$~eV$^{-1}$ along the symmetric directions. The red solid line (squares) corresponds to the to the result from the orbital-summed susceptibilities, given by Eq. (\ref{JqAvDef}); the violet dot-dashed line (circles) corresponds to the orbital averaged echange interaction from the orbital-resolved inverse susceptibilities.
  The short-dashed blue line with triangles shows the result of DFT approach of Ref. \onlinecite{DFTJ}. 
  The dashed line without symbols shows the momentum dependence of the nearest-neighbor exchange. The inset shows Brillouin zone with the symmetric points.}
\label{Fig_Jq_fcc}
\end{figure}

  \begin{figure}[t]
		\center{		\includegraphics[width=0.8\linewidth]{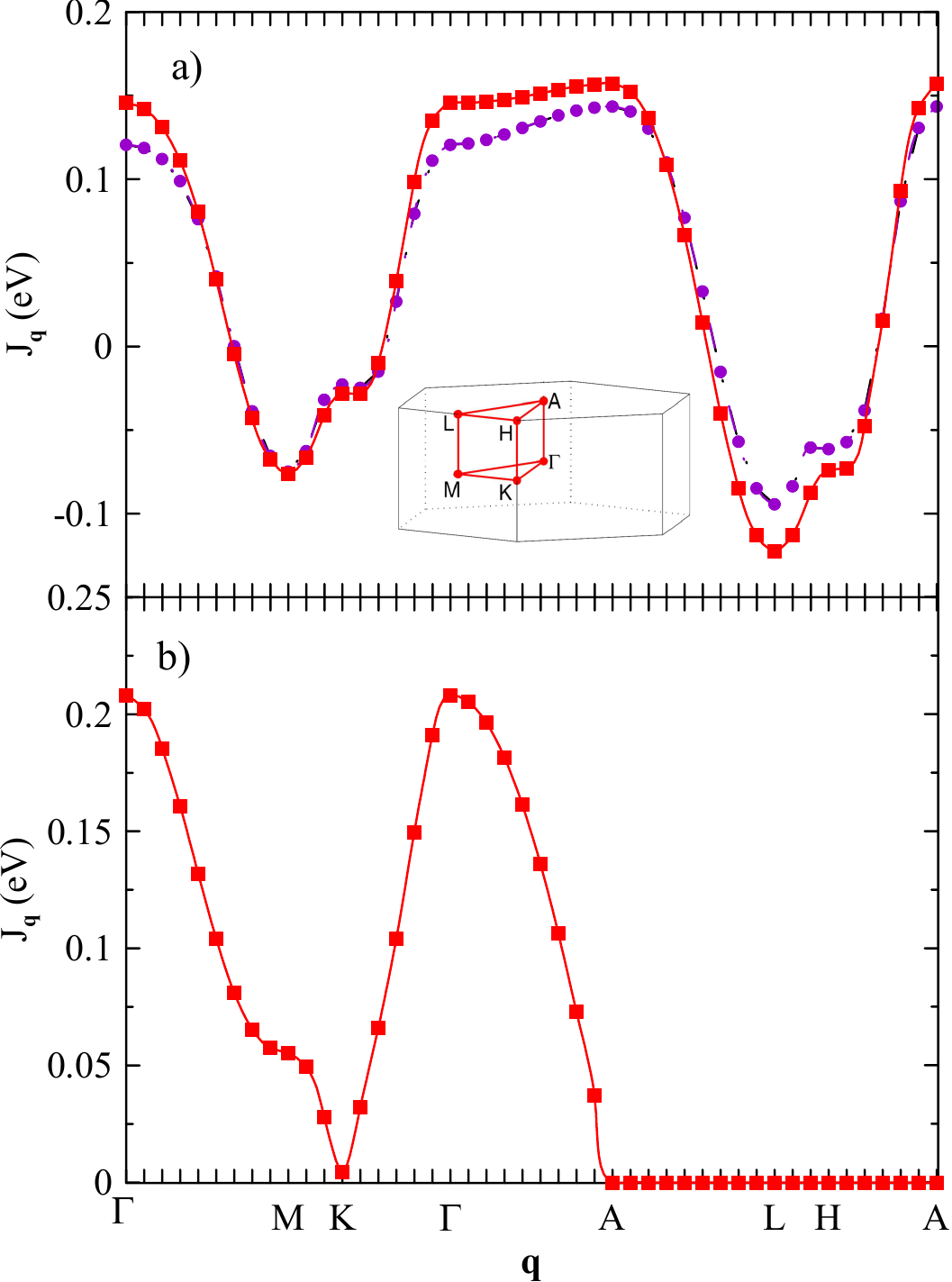}}
		\caption{Momentum dependence of the exchange interactions $J^{11}_{\bf q}$ (a) and $|J^{12}_{\bf q}|$ (b) in hcp cobalt at $\beta=7$~eV$^{-1}$ along the symmetric directions. The notations are the same as in Fig. \ref{Fig_Jq_fcc}. The inset shows Brillouin zone with the symmetric points.}
\label{Fig_Jq_hcp}
\end{figure}

\begin{figure}[b]
		\center{		\includegraphics[width=0.95\linewidth]{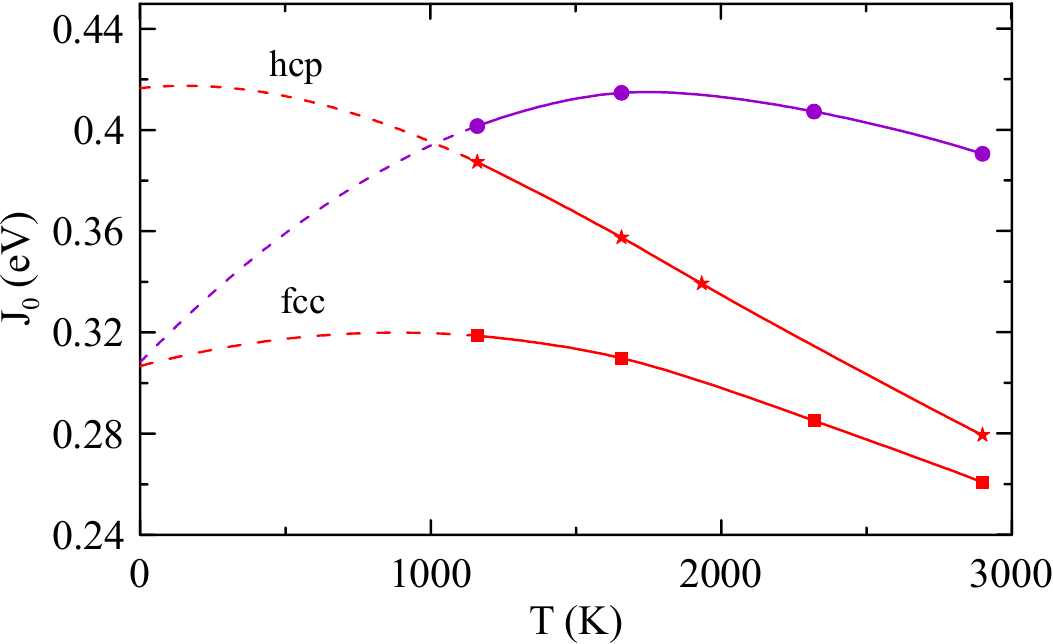}}
		\caption{Temperature dependence of the exchange interaction $J_0$ in fcc cobalt (solid line with squares for the orbital summed inverse susceptibility and solid line with circles for the interaction from the orbital resolved susceptibilities) and $J^{11}_0+J_0^{12}$ in hcp cobalt (solid line with stars). Dashed lines show the result of extrapolation.}
\label{Fig_JT}
\end{figure}

We first consider exchange interactions in fcc phase. Figure~{\ref{Fig_Jq_fcc}} shows the momentum dependence of the obtained exchange interaction at $\beta=7$~eV$^{-1}$ near the DMFT Curie temperature $T_{\rm C}$. Similarly to the previous study of nickel \cite{MyJ}, Eq.~(\ref{JqAvDef}), considering the orbital-summed susceptibilities yields smaller exchange interaction than the average of the orbital-resolved exchange interactions, although the difference between various approaches is not as large as in nickel, which is related to the presence of well-defined local magnetic moments in Co. In the following we consider mainly the exchange interactions from the orbital summed susceptibilities (\ref{JqAvDef}), since it produced most reasonable results for iron and nickel\cite{MyJ}. {The obtained momentum dependence of the exchange interaction near the $\Gamma$ point is well described by the 
nearest-neighbor exchange interaction, although  somewhat deviates from it near X and L points of the Brillouin zone. For comparison we also show the result of the DFT approach of Ref. \onlinecite{DFTJ}. One can see that the DFT approach overestimates the exchange interaction near the $\Gamma$ point (see also the comparison of the magnon dispersion with the experimental data below in Sect. \ref{MagDisp}).

 Figure~{\ref{Fig_Jq_hcp}} shows the momentum dependencies of the obtained exchange interactions at $\beta=7$~eV$^{-1}$ in hcp cobalt near the DMFT Curie temperature $T_{\rm C}$. One can see that the exchange interaction at zero momentum $J_0^{11}+J_0^{12}$ is somewhat larger than the interaction $J_0$ in the fcc phase, which provides larger Curie temperature. Since the intersublattice interaction $J_{\bf q}^{12}\propto \cos((2n+1)q_z d)$ with integer $n$ and the distance $d=c/2$ between planes of cobalt atoms, corresponding to different sublattices,  the exchange interaction $J_{\bf q}^{12}$ vanishes at the upper (lower) edge of the Brillouin zone ($q_z=\pm \pi/c$).

The temperature dependencies of exchange interactions are shown in Fig. \ref{Fig_JT}. At low temperatures the obtained exchange interaction, determined from the inverse magnetic susceptibility, $J_0\simeq 0.3$~eV for fcc structure ($J^{11}_0+J^{12}_0\simeq 0.4$~eV  for the hcp structure), is relatively weakly temperature dependent. Near the Curie temperature it is twice (more than twice for hcp structure) larger than the corresponding value for iron \cite{MyJ}. However, the local magnetic moment of cobalt ($p_{\rm loc}\simeq 1$) is smaller than that of iron ($p_{\rm loc}\simeq 1.5$), see Ref. \onlinecite{MyJ}.  Considering also the correction $T_W=\sqrt{2}T_K$ in the mean-field equation $T_C=J_0 p_{\rm loc}(p_{\rm loc}+1)/3-T_W$, which appears from the paramagnetic Weiss temperature of local spin susceptibility (\ref{chiloc}), related to the local magnetic moment screening, the Curie temperature of cobalt in DMFT is larger than the experimental Curie temperature of iron by only 1.4 (1.6) times in the fcc (hcp) phase. Therefore, despite relatively large exchange interaction, the Curie temperature of cobalt is suppressed because of smaller local magnetic moment and its Kondo screening. We note, that the extension of the presented approach to the symmetry broken phase is necessary to estimate accurately the exchange interactions in the low-temperature regime, which is beyond the scope of the present paper.

\subsection{Magnon dispersions and spin-wave stiffnesses}\label{MagDisp}

Using the obtained exchange interactions in the temperature range $T\gtrsim T_C$, it is useful to consider extrapolation of the obtained results to the low-temperature region, assuming that the exchange interactions do not change strongly with lowering the temperature. In Fig. \ref{FigD_fcc}, we show the comparison of the experimental magnon dispersion \cite{Dfcc} in the fcc phase of cobalt at room temperature (FeCo alloy containing 8\% of iron was used to stabilize the crystal structure) to the dispersion obtained from the exchange interactions as $E_{\mathbf q}=p_{\rm loc}(J_0-J_{\mathbf q})$ (the exchange interactions from the orbital-summed susceptibilities are considered). One can see that the magnon dispersion agrees well with the experimental data. The respective spin wave stiffness $D=290$~meV$\cdot\AA^2$ is somewhat smaller than the experimental data ($D=360$~meV$\cdot\AA^2$), which can be attributed to the larger considered temperature. On the other hand, the DFT approach of Ref. \onlinecite{DFTJ} yields much larger spin wave stiffness $D=663$~meV$\cdot\AA^2$.

\begin{figure}[b]
		\center{		\includegraphics[width=0.93\linewidth]{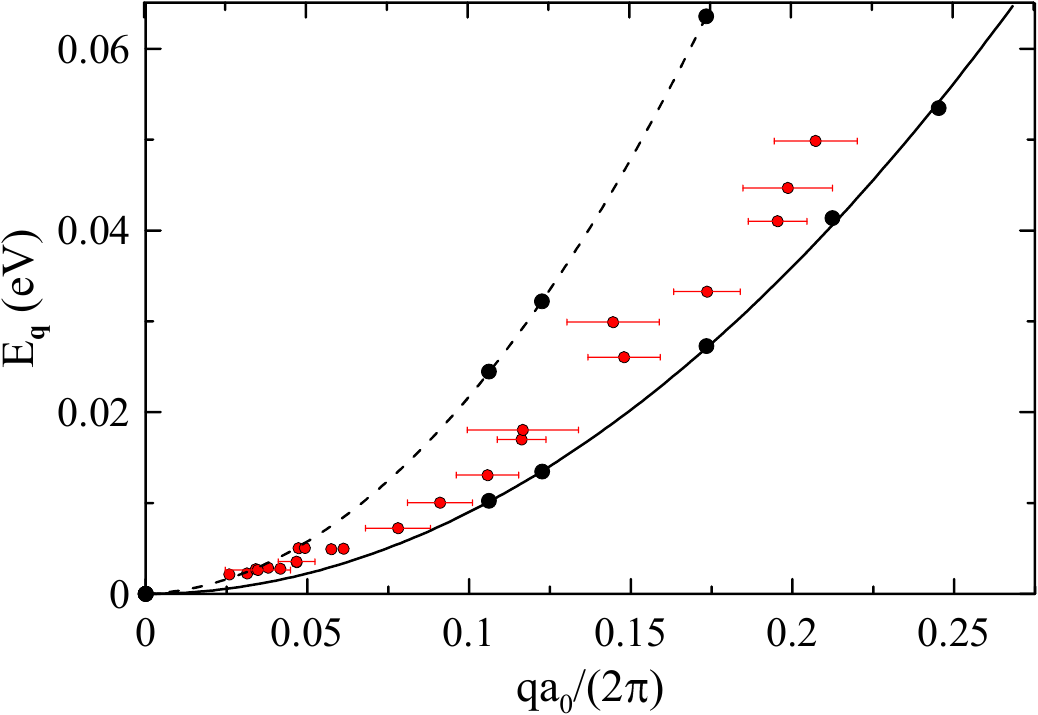}}
		\caption{Comparison of the experimental magnon dispersion in fcc cobalt along the [111] direction (Ref. \onlinecite{Dfcc}) at room temperature (symbols) with the results of DFT+DMFT approach at $\beta=7$~eV$^{-1}$ (solid line) and the DFT approach of Ref. \onlinecite{DFTJ} (dashed line).}
\label{FigD_fcc}
\end{figure}

 \begin{figure}[t]
		\center{		\includegraphics[width=0.95\linewidth]{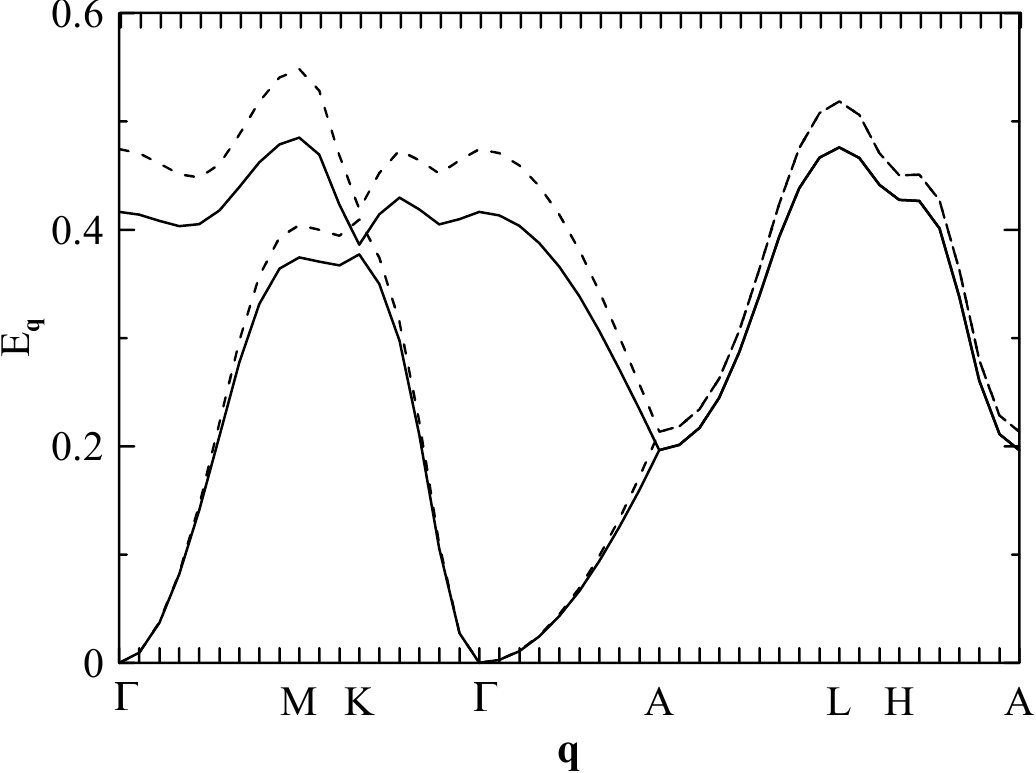}}
		\caption{Magnon dispersion in hcp cobalt at $\beta=7$~eV$^{-1}$ (solid lines) and $\beta=10$~eV$^{-1}$ (dashed lines).}
\label{Disp_hcp}
\end{figure}

The magnon dispersion in hcp phase is obtained as the ${\bf q}$-dependent eigenvalues of the matrix of the spin-wave Hamiltonian
\begin{equation}
    H_{\rm SW}({\mathbf q})=p_{\rm loc}\left(\begin{array}{cc}
        J_0^{11} + 
        J_0^{12}
- J_{\mathbf{q}}^{11} & - J_{\mathbf{q}}^{12} \\
        - J_{\mathbf{q}}^{21} & J_0^{22} + 
        J_0^{21} 
        - J_{\mathbf{q}}^{22} 
    \end{array}\right)
\end{equation}
and shown in Fig. \ref{Disp_hcp}. In Fig. \ref{FigD_hcp}, we show the comparison of the experimental magnon dispersion at room temperature \cite{Dhcp} to the low-energy part obtained from the exchange interactions extracted from the orbital-summed susceptibilities. One can see that the magnon dispersion agrees well with the experimental data. The respective spin wave stiffness $D=405$~meV$\cdot\AA^2$ is larger than in the fcc phase, but also remains smaller than the experimental value ($D=490$~meV$\cdot\AA^2$) because of larger considered temperature. The reasonable agreement presented above shows correctness of the obtained exchange interactions, which are used in the next subsection to estimate the non-local corrections to the Curie temperature.

\begin{figure}[t]
		\center{		\includegraphics[width=0.89\linewidth]{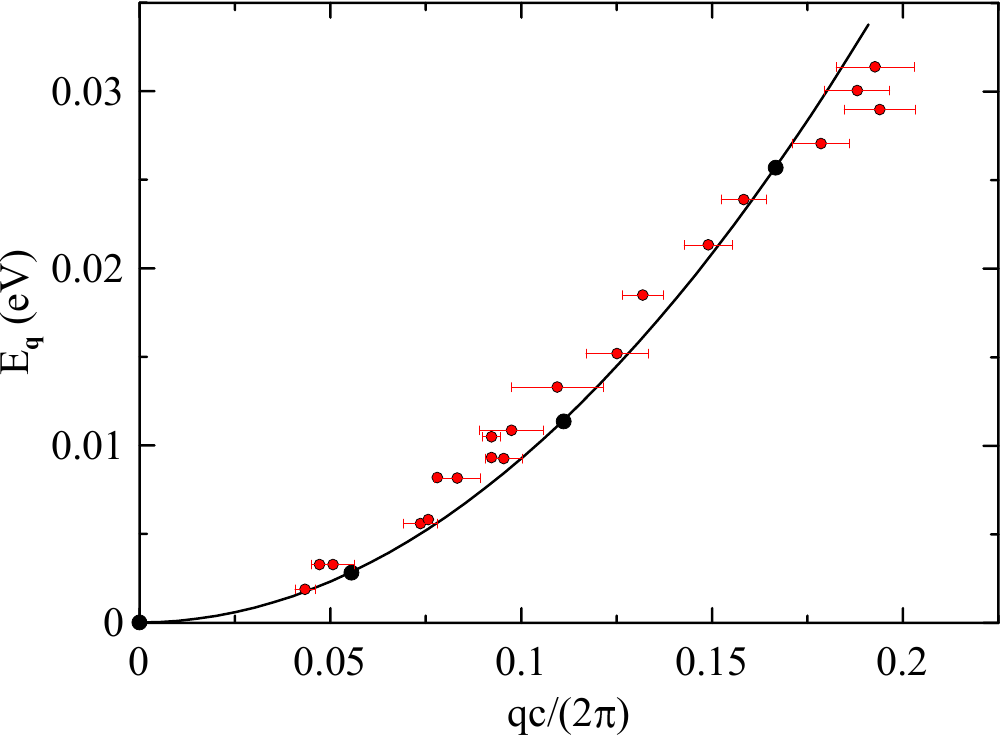}}
		\caption{Comparison of the experimental magnon dispersion in hcp cobalt along the [001] direction (Ref. \onlinecite{Dhcp}) at room temperature (symbols) with the results of DFT+DMFT approach at $\beta=10$~eV$^{-1}$ (solid line).}
\label{FigD_hcp}
\end{figure}

\subsection{Curie temperatures in the spherical model and implication of the obtained results for hcp-fcc transformations}
\label{SW}

To estimate the Curie temperatures of the fcc and hcp cobalt beyond DMFT, we consider the effective spherical model \cite{Berlin,Baxter,Nagaev,OurFe_ag}
\begin{equation}
\mathcal{S}=\frac{1}{2T}\sum_{{\mathbf q},rr'} \left[\lambda_r |{\mathbf S}_{\mathbf q}^r|^2 \delta_{rr'}-J^{rr'}_{\mathbf q} {\mathbf S}_{\mathbf q}^r {\mathbf S}_{-\mathbf q}^{r'}\right].
\label{sr}
\end{equation}
Assuming that the sites of the unit cell are equivalent, the constant $\lambda_r=\lambda$ is determined in paramagnetic phase by the sum rule
\begin{equation}
\sum_{\mathbf q}\langle |{\mathbf S}_{\mathbf q}^r|^2 \rangle=\frac{3T}{2}\sum_{\mathbf q}\left[\lambda \delta_{rr'}-J_{\bf q}\right]_{rr}^{-1}=T\chi_{\rm loc}.
\end{equation}
The Curie temperature is determined by the zero of the lowest eigenvalue of the matrix $\lambda \delta_{rr'}-J_{\mathbf q}$ at ${\mathbf q}=0$, which yields $\lambda=\sum_{r'} J^{rr'}_0$. Using the temperature dependence of the local susceptibility (\ref{chiloc}), we find
\begin{equation}
T_C=
\frac{p_{\rm loc}(p_{\rm loc}+1)}
{3\sum\limits_{{\mathbf q}} \left[\lambda \delta_{rr'}-J_{\mathbf q}^{rr'}\right]_{11}^{-1}}-T_W.
\end{equation}
For $T_W=0$ this result coincides with the result of RPA approach \cite{RPA_TC} for equivalent atoms in the unit cell. Taking the obtained exchange interactions at $\beta=10$~eV$^{-1}$ we obtain $T_C\simeq 740$~K for the fcc phase and $T_C\simeq 1250$~K for the hcp phase. Therefore, with account of the non-local corrections, the Curie temperature of the hcp phase remains larger than that of the fcc phase. 

We note that the DFT energy of magnetic hcp phase is lower than that of the fcc phase by only $\sim 0.02$~eV/atom. At the same time, due to magnetic order these phases gain the energy $\sim 0.2$~eV/atom in the ground state, see Ref. \onlinecite{EnergiesNew}. Therefore, loss of the long-range magnetic order in the fcc phase would make it energetically unfavourable in comparison to the (partly) ferromagnetically ordered hcp phase. This can lead to the reappearance of hcp phase at higher temperatures in the vicinity of Curie temperature.

To estimate the effect of the local and spin correlations, apart from the DFT energy of non-magnetic state $E_{\rm DFT}$, we also consider the potential energy $E_{\rm pot}$ of the on-site Coulomb repulsion, determined within CT-QMC and subtracted by the double counting contribution
\begin{equation}
E_{\rm dc}=\langle {n}_{ird} \rangle^2  \left[U(2n_{\rm orb-1})-J(n_{\rm orb}-1)\right]/(4n_{\rm orb}),
\end{equation}
the spin fluctuation contribution per atom
\begin{equation}
E_{\rm sfl}=-\frac{3}{2}\sum\limits_{{\mathbf q},r} J^{1r}_{\mathbf q}\left[\lambda \delta_{rr'}-J_{\mathbf q}^{rr'}\right]_{r1}^{-1},
\end{equation}
and, finally, the magnetic contribution $E_m=-\sum_{r} J^{1r}_{\mathbf 0} \bar{S}^2/2$. To determine the magnetization $\bar{S}$ per atom, we generalize the sum rule (\ref{sr}) to ferromagnetic phase, \begin{align}
\alpha \bar{S}^2&+\sum_{\mathbf q}\langle |{\mathbf S}_{\mathbf q}^r|^2 \rangle\notag\\&=\alpha \bar{S}^2+\frac{3T}{2}\sum_{\mathbf q}\left[\lambda \delta_{rr'}-J_{\bf q}\right]_{rr}^{-1}=T\chi_{\rm loc},
\end{align}
where we choose $\alpha=T/(T+T_W)$ to reflect the decrease of the local magnetic moment due to finiteness of the Weiss temperature. This yields 
\begin{equation}
\bar{S}^2=\frac{\mu_{\rm loc}^2}{4\mu_B^2}\frac{T_C-T}{T_C+T_W}.
\end{equation}

\begin{table}
\begin{tabular}{||l|l|l|l|l|l||}
\hline\hline
& $E_{\text{DFT}}-E_{\text{DFT}}^{\text{hcp}}$ & $E_{\rm pot}-E_{\rm dc}$ & $E_{%
\text{sfl}}$ & $E_{m}$ & $E_{\rm tot}$ \\ \hline
fcc & $-0.02$ & $-2.47$ & $-0.06$ & $0$ & $-2.55$ \\ \hline
hcp & $0$ & $-2.47$ & $-0.04$ & $-0.07$ & $-2.58$ \\ \hline
fcc-hcp & $-0.02$ & $0$ & $-0.02$ & $+0.07$ & $+0.03$ \\ \hline\hline
\end{tabular}
\caption{Various contributions to energies of fcc and hcp phases of cobalt at $\beta=10$~eV$^{-1}$ (in units of eV/atom) together with the total energy (relative to the $E_{\rm DFT}^{\rm hcp}$).}
\end{table}

The above specified contributions to the total energy are listed for $\beta=10$~eV$^{-1}$ in Table I. One can see that while the contributions $E_{\rm pot}-E_{\rm dc}$, corresponding to the effect of local correlations, are close for both phases, and the spin fluctuation contribution leads to the gain of the energy of fcc phase in comparison to the hcp one due to smaller exchange interaction, the DFT and $E_{\rm sfl}$ contributions are compensated by the energy gain of hcp phase due to its finite magnetization, and therefore hcp phase becomes more energetically favourable near Curie temperature.

As we discuss above, the drop of magnetization, which is accompanied by reducing the content of the fcc phase at the temperature $T\sim 700$~K, was observed recently in cobalt in Ref. \onlinecite{PAC}. At that temperature the reappearance of hcp phase was obtained, albeit with suppressed magnetization. In this light, higher Curie temperature of hcp phase may correspond to the one, observed in the experiment.

}

\section{Conclusion}
In summary, we have evaluated local magnetic moments and Curie temperatures in the fcc and hcp phases of cobalt within DFT+DMFT approach. The hcp phase has larger Curie temperature. 
To obtain non-local corrections to Curie temperatures, we have evaluated exchange interactions using recently proposed approach in the paramagnetic phase. 
We show that larger Curie temperature of hcp phase originates from larger exchange interaction 
($J_0\simeq 0.3$~eV in fcc phase 
and $J_0\simeq 0.4$~eV in hcp phase). 
This approach also allows a correct description of the experimental data for the spin-wave stiffness. The obtained magnon dispersions are in reasonable agreement with the experimental data. 

Using obtained exchange interactions we have estimated non-local corrections to Curie temperatures. We show that with account of non-local corrections the Curie temperature in the fcc phase ($T_{C,{\rm fcc}}\simeq 740$~K) remains smaller than that of the hcp phase ($T_{C,{\rm hcp}}\simeq 1250$~K). Therefore, we expect that because of the loss of magnetic energy the fcc phase becomes unstable at the temperatures close to $T_{C,{\rm hcp}}$. We confirm this conjecture by explicit calculation of various contributions to the energies of fcc and hcp phases at $\beta=10$~eV$^{-1}$.

Further experimental studies of crystal structure of cobalt near Curie temperature can provide more information on the possibility of existence of hcp phase near Curie temperature. The proposed method can be applied to the other substances experiencing structural transition which originate from the loss of magnetic order. Also, it can be used for calculation of non-local corrections to magnetic transition temperatures. 

Accurate estimate of the entropy contribution to the free energy of cobalt is of certain interest. Also, extension of the considered approach to SU(2) form of electron interaction represents important field of future development.

{\it Acknowledgements}. The author is grateful to I. A. Goremykin for discussions and the help with using Wannier90 package. Performing the 
DMFT calculations was supported by the Russian Science Foundation
(project 19-72-30043-P). The DFT calculations are supported by the theme ``Quant" 122021000038-7 of {the} Ministry of Science and Higher Education of the Russian Federation. The calculations were performed on the cluster of the Laboratory of Material Computer Design of MIPT and the Uran supercomputer at the IMM UB RAS.

\appendix





\appendix
\section*{Appendix}

\begin{figure}[h!]
		\center{
		\includegraphics[width=0.9\linewidth]{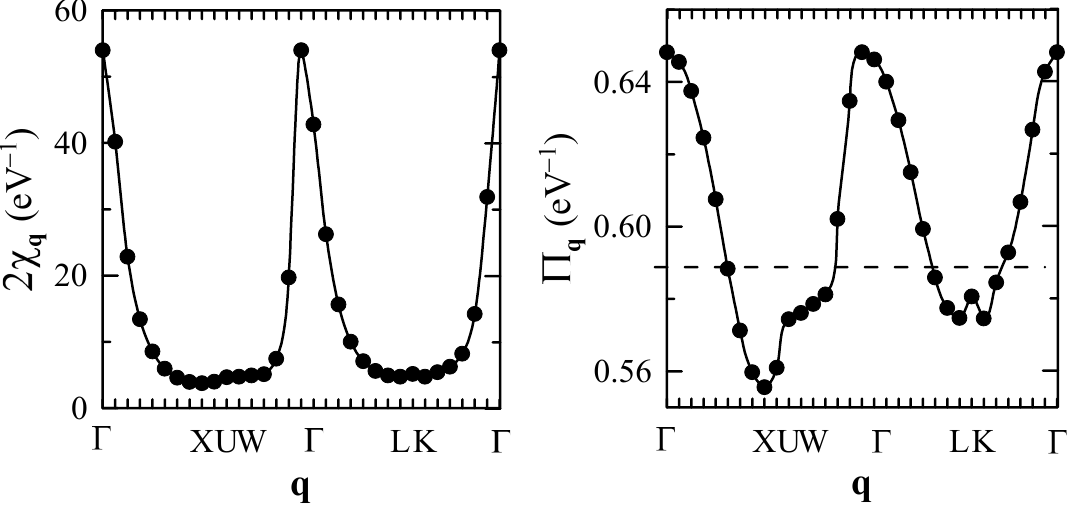}}
		\caption{
		Momentum dependence of the orbital-summed non-local susceptibility $\chi_{\bf q}$, and its particle-hole irreducible counterpart $\Pi_{\bf q}$ in fcc cobalt at $\beta=7$~eV$^{-1}$ along the symmetric directions.}
		 \label{Fig_chiq_Co}
\end{figure}

\begin{figure}[h!]
		\center{
		\includegraphics[width=0.9\linewidth]{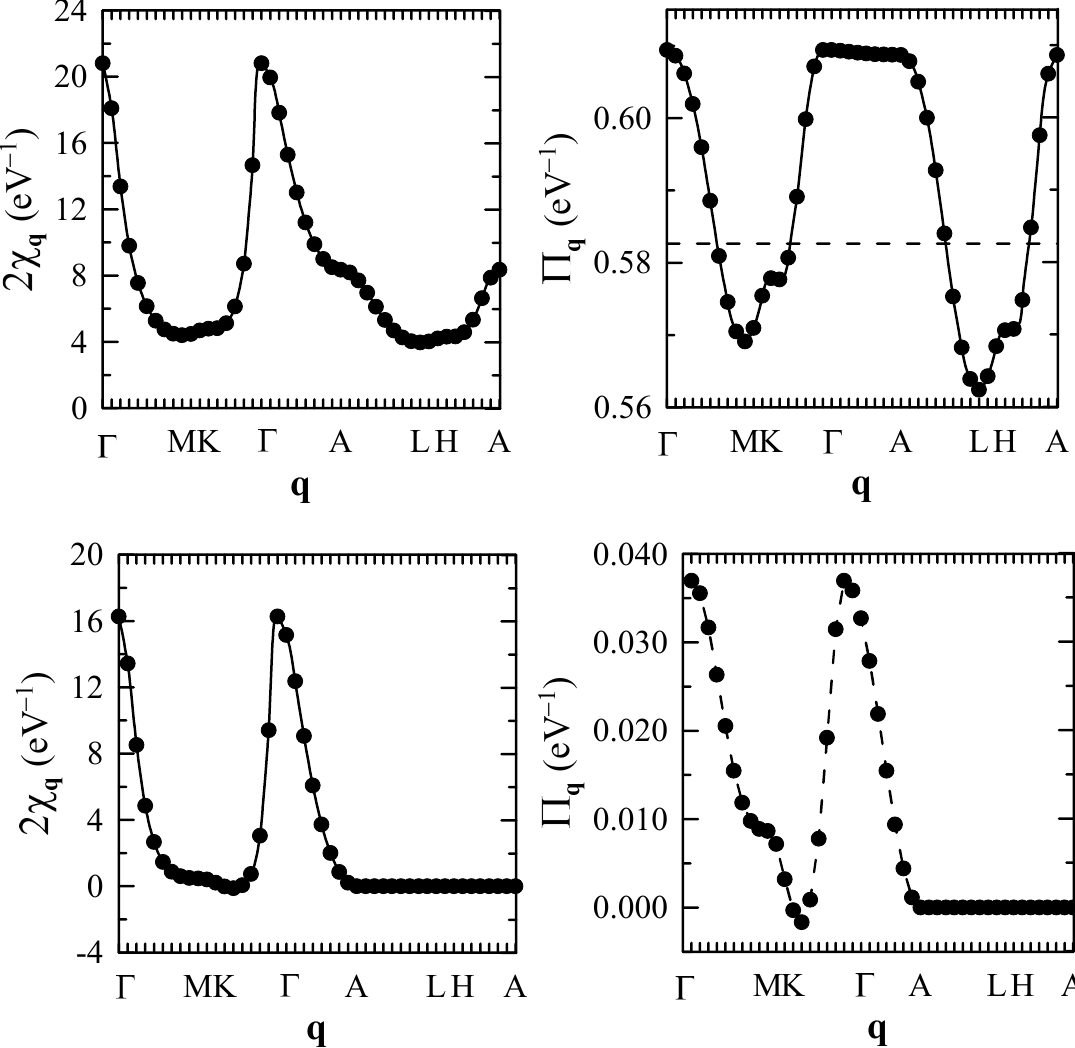}}
		\caption{
		Momentum dependence of the orbital-summed non-local susceptibility $\chi_{\bf q}^{rr'}$, and its particle-hole irreducible counterpart $\Pi^{rr'}_{\bf q}$ in hcp cobalt at $\beta=6$~eV$^{-1}$ along the symmetric directions. Top row: $r,r'=1$ component, lower row $r=1,r'=2$.}
		 \label{Fig_chiq_Co1}
\end{figure}

In this Appendix we present the momentum dependence of the non-local susceptibility of fcc cobalt at $\beta=7$~eV$^{-1}$ (Fig. \ref{Fig_chiq_Co}) and hcp cobalt at at $\beta=6$~eV$^{-1}$ (Fig. \ref{Fig_chiq_Co1}), together with the orbital-summed polarization operators $\Pi_{\bf q}$. One can see that the susceptibilities are strongly peaked at $q=\Gamma=0$ showing strong ferromagnetic correlations. 

\end{document}